\documentclass[1pt,letterpaper]{article}
\usepackage[margin=1 in]{geometry}
\usepackage[latin1]{inputenc}
\usepackage{amsmath,amsthm}
\usepackage{amsmath,bm}
\usepackage{amsmath,mathtools}
\usepackage{amsfonts}
\usepackage{amssymb}
\usepackage{eurosym}
\usepackage{float}
\usepackage{esint}
\usepackage{xcolor}
\usepackage{titling}
\usepackage{graphicx}
\usepackage{caption}
\usepackage{subfigure}
\usepackage{hhline}
\usepackage{cite}
\usepackage[section]{placeins}
\usepackage[pdftex,colorlinks=true,urlcolor=blue,citecolor=brown]{hyperref}

\theoremstyle{remark}

\theoremstyle{definition}

\numberwithin{equation}{section}

\begin{document}

\title{The effects of network architecture on the photomechanical
performance of azo-acrylate liquid crystal elastomers}
\author{Anastasiia Svanidze$^1$, Sudarshan Kundu$^1$, Olena Iadlovska$^1$,
\and Anil K.Thakur$^{1}$, Xiaoyu Zheng$^{2}$ and Peter Palffy-Muhoray$^{1,2}$
\\
\emph{$^1$Advanced Materials and Liquid Crystal Institute, Kent State
University, OH, USA}\\
\emph{$^2$Department of Mathematical Sciences, Kent State University, OH,
USA } }
\maketitle

\begin{abstract}
Azo-containing liquid crystal elatomers are photomechanical materials which
can be actuated by illumination. The photomechanical response is a result of
the photoisomerization of the azo moiety, which produces bulk stresses in
the material. These stresses arise via two distinct and competing
mechanisms: order parameter change induced stress and direct contractile
stress. We describe thermomechanical and photomechanical experiments aimed
at assessing the relative contributions of these. we discuss our results and
summarize our findings.
\end{abstract}

\section{Introduction}

\bigskip

Liquid crystal elastomers (LCEs) are a unique class of advanced materials
that combine properties of liquid crystals and rubber-like materials. They
consist of a crosslinked polymer network polymerized with monomers
possessing liquid crystalline phases. LCEs, first proposed by P.G. de Gennes 
\cite{PGdG} and realized by H. Finkelmann \cite{Heino}, are remarkable
materials \cite{ppm1} due to their exceptional responsiveness to external
stimuli such as heating or illumination. This responsiveness originates from
the sensitivity of liquid crystalline order to system parameters near a
phase transition. LCEs hold significant potential for applications, as they
can sustain and exert shear stresses as elastic solids. This makes them
highly versatile for various applications requiring shape changes and stress
response, such as soft robotics \cite{robot1, robot2, Broer4D18}, artificial
muscles \cite{muscle1, muscle2}, adaptive optics \cite{opt1}, smart textiles 
\cite{opt2, textile1, textile2} as well as biomedical applications \cite%
{bio1, bio2, bio3}. The combination of mechanical flexibility and tunable
properties makes LCEs well suited for next-generation responsive and
adaptive technologies.

A great variety of LCEs has been synthesized, ranging from Finkelmann's
early polyhydrosiloxane-based ones  \cite{Heino, CviklinskiTerentjev02,
HoganTerentjev02, HarveyTerentjev07, SanchezHeino11, DawsonPalffy11,
JavierDASA24} to more recent acrylate-based \cite{Broer4D18, WhitePolymer15,
WhiteScience15, WhiteAzoLCE16, Ware4Dprinting17, White3Dprint18, Broer4D20,
5 elastic, regimes, White1, Gleeson24, Yakacki15}, polyester-based materials 
\cite{polyester1, polyester2} and others \cite{other1, other2}. Actuation
mechanisms used to change the liquid crystal order parameter include heating 
\cite{Broer4D18, muscle1, bio2,WhitePolymer15, WhiteScience15,
Ware4Dprinting17, White3Dprint18}, photoactuation  \cite{ppm1, robot1,
CviklinskiTerentjev02, HoganTerentjev02, HarveyTerentjev07, SanchezHeino11,
DawsonPalffy11, JavierDASA24, WhiteAzoLCE16, Broer4D20, regimes, White1,
WarnerB, ppm2, ppm3, ppm4, Warner04, Warner06, Warner09}, the application of
electric \cite{electric1, electric2, electric3} and magnetic \cite{magnetic1}
fields and chemical stimuli \cite{chem1}.

In LCEs, photoactuation can be achieved by incorporating azobenzene moieties
into the polymeric network. Upon illumination, the azobenzene undergoes
photoisomerization, inducing internal stresses that can result in shape
changes and enable the material to perform mechanical work. The
photomechanical response can be attributed to two competing mechanisms:
changes in the liquid crystal orientationqal order parameter, and direct
contractile stress exerted on the network. In this work, we focus on
amine-acrylate LCEs doped with azo dyes with different degrees of attachment
to the network. We have measured and compared the thermal and photoresponse
of three types of samples: those in which the azobenzene moiety is
covalently bonded into the network at both ends (2-azo), at one end (1-azo)
and not bonded at all (free-azo). In this paper, we describe our sample
materials, report experimental results, discuss our findings in light of the
two competing mechanisms and conclude by briefly summarizing our results.

\section{Sample Composition and Structure}

\subsection{Materials}

\begin{figure}[h]
\centering
\includegraphics[width=0.8\linewidth]{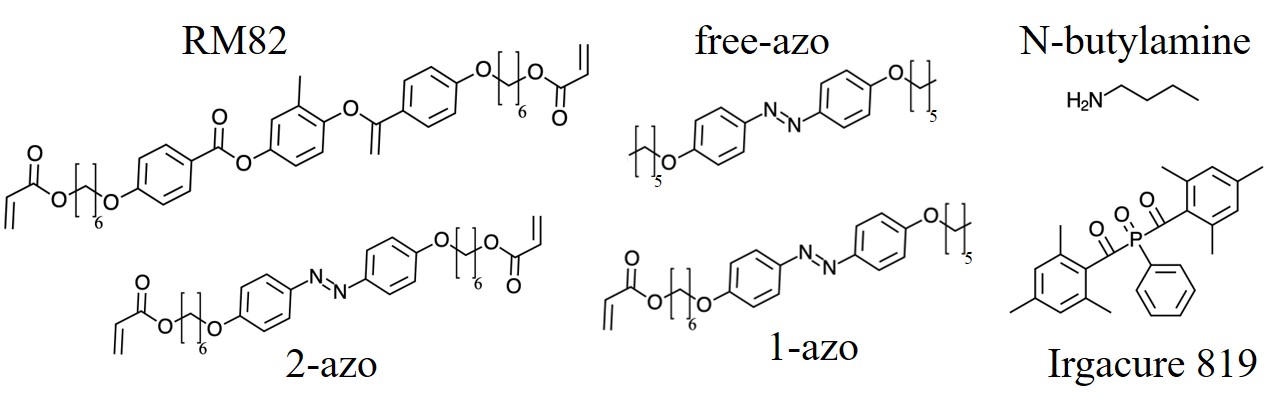}
\caption{Chemical constituents of 2-azo, 1-azo and free-azo samples}
\label{fig:chemistry}
\end{figure}

The chemical structure of the constituents of our samples are shown in Fig.~%
\ref{fig:chemistry}. The liquid crystal monomer
1,4-Bis[4-(6-acryloyloxyhexyloxy)benzoyloxy]-2-methylbenzene (RM82) was
purchased from Jiangsu Hecheng Advanced Materials Co., Ltd (China). The azo
dyes 4,4'-Bis(6-acryloyloxyhexyloxy)azobenzene (2-azo) and acrylic acid
(4-((4-hexyloxyphenyl) diazenyl)phenoxy)hexyl ester (1-azo) were purchased
from SYNTHON Chemicals GmbH \& Co. KG (Germany). The azo dye
1,2-Bis(4-(hexyloxy)phenyl)diazene (free-azo) was purchased, as result of
custom synthesis, from Henan Daken Chemical Co., ltd (China). \ The chain
extender N-butylamine and the photoinitiator
phenylbis(2,4,6-trimethylbenzoyl)phosphine oxide (Irgacure 819) were
purchased from Sigma-Aldrich, Merck KGaA (Germany). All materials were used
as received; monomer purity was not taken into account in determining molar
ratio. Polyimide SE-2170 (Nissan) and its thinner 21 (Nissan) for liquid
crystal alignment were purchased from Brewer Science, Inc. (USA).

Amine-acrylate-based LCEs were first proposed by T.J. White \textit{et.\
al.\ }\cite{WhiteScience15}; a number of the recipes are available for
fabrication today \cite{WhitePolymer15, WhiteScience15, WhiteAzoLCE16,
Ware4Dprinting17, White3Dprint18, Broer4D18, Broer4D20}. Below we briefly
describe our fabrication process.

\subsection{Network synthesis and sample production}


Plain glass substrates, $10~{cm}\times 10~{cm\times }1~{mm}$, were cleaned
in ultrasonic bath with detergent at $60^{\circ }\mathrm{C}$ for $15~{min}$
followed by rinsing of substrates with DI water and IPA. Following this,
substrates were dried in an oven for $15~{min}$ at $82^{\circ }\mathrm{C}$.
The polyimide mixture SE-2170, in proportion $1:3$ with its thinner, was
spin coated onto the clean substrates with the sequence: $1~{s}$ $-500~rpm$, 
$30~{s}$ $-1500~rpm$, $1~{s}-50~rpm$. After spin coating, the polyimide was
soft baked at $80^{\circ }\mathrm{C}$ on the hotplate for $5~{min}$ and
then hard baked in the oven at $200^{\circ }\mathrm{C}$ for $45~{min}$. The
glass substrates were then cooled to room temperature, and were rubbed
unidirectionally with velvet cloth on a rubbing block $10-15$ times. They
were subsequently cut into $4.5~cm\times 4.5$~$cm$ squares and assembled
anti-parallel into cells using Mylar film (DuPont Teijin Films) as spacers
and Norland Optical Adhesive 68T as glue.

The liquid crystal monomer, azo dye and photoinitiator were added to the
amine and melted at $110^{\circ }\mathrm{C}$ while stirring for $30$~$s$.
The mass of the photoinitator added was $2.5~wt.\%$ of the combined mass
of the other constituents. Relative concentrations of the components of six
samples are shown in Table \ref{table_1}.

\begin{table}[h]
\caption{Compositions of sample types}
\label{table_1}\centering
\begin{tabular}{|c||c|c|c|c|}
\hline
sample type & RM82 & azo dye & acryl./amine & Irgacure \\ 
& (mole fraction) & (mole fraction) & (mole ratio) & (wt.~\%) \\ \hline
no azo & 1 & 0 & 1.1:1 & 2.5 \\ \hline
free-azo 2\% & 0.98 & 0.02 & 1.1:1 & 2.5 \\ \hline
1-azo 2\% & 0.98 & 0.02 & 1.1:1 & 2.5 \\ \hline
2-azo 2\% & 0.98 & 0.02 & 1.1:1 & 2.5 \\ \hline
free-azo 5\% & 0.95 & 0.05 & 1.1:1 & 2.5 \\ \hline
1-azo 5\% & 0.95 & 0.05 & 1.1:1 & 2.5 \\ \hline
2-azo 5\% & 0.95 & 0.05 & 1.1:1 & 2.5 \\ \hline
\end{tabular}%
\end{table}

Empty cells with planar surface alignment on the hotplate at $85^{\circ }%
\mathrm{C}$ were filled with the mixture via the capillary effect. Filled
cells were moved to an oven at $77^{\circ }\mathrm{C}$ and left overnight
for oligomerization.

The LCE cells were cured the following day using a mercury lamp (OmniCure
s1500, Excelitas Technologies Corp. USA). A filter with cut-on wavelength $%
400~nm$ (20CGA-400, Newport, USA) was placed on each cell to reduce the
photoisomerization of azobenzene during curing. Polymerization was carried
out with light intensity of $75$~${mW}/{cm}^{2}$ at $70^{\circ }\mathrm{C}$
for $20$~$min$. Light intensity was measured at $405$~$nm$ with Optical
Power Meter (1830-R, Newport, USA).

After curing, cells were heated to $100^{\circ }\mathrm{C}$ and opened
using a thin blade. The LCE samples were removed from the substrate with
tweezers. 

The differences between samples are as follows: 2-azo samples contained azo
molecules with two covalently bondable double bonds, 1-azo samples contained
azo molecules with one covalently bondable double bond and free-azo samples
contained azo molecules with no covalentlly bondable double bonds. All
samples were produced using identical procedures, only the type (2-azo,
1-azo and free-azo) and percentage ($2\%$ or $5\%$) of azo compounds was
changed.

A simple schematic of the resulting polymer network, shown in Fig.~\ref%
{fig:network}, illustrates the key structural differences.

\begin{figure}[h]
\centering
\includegraphics[width=1\linewidth]{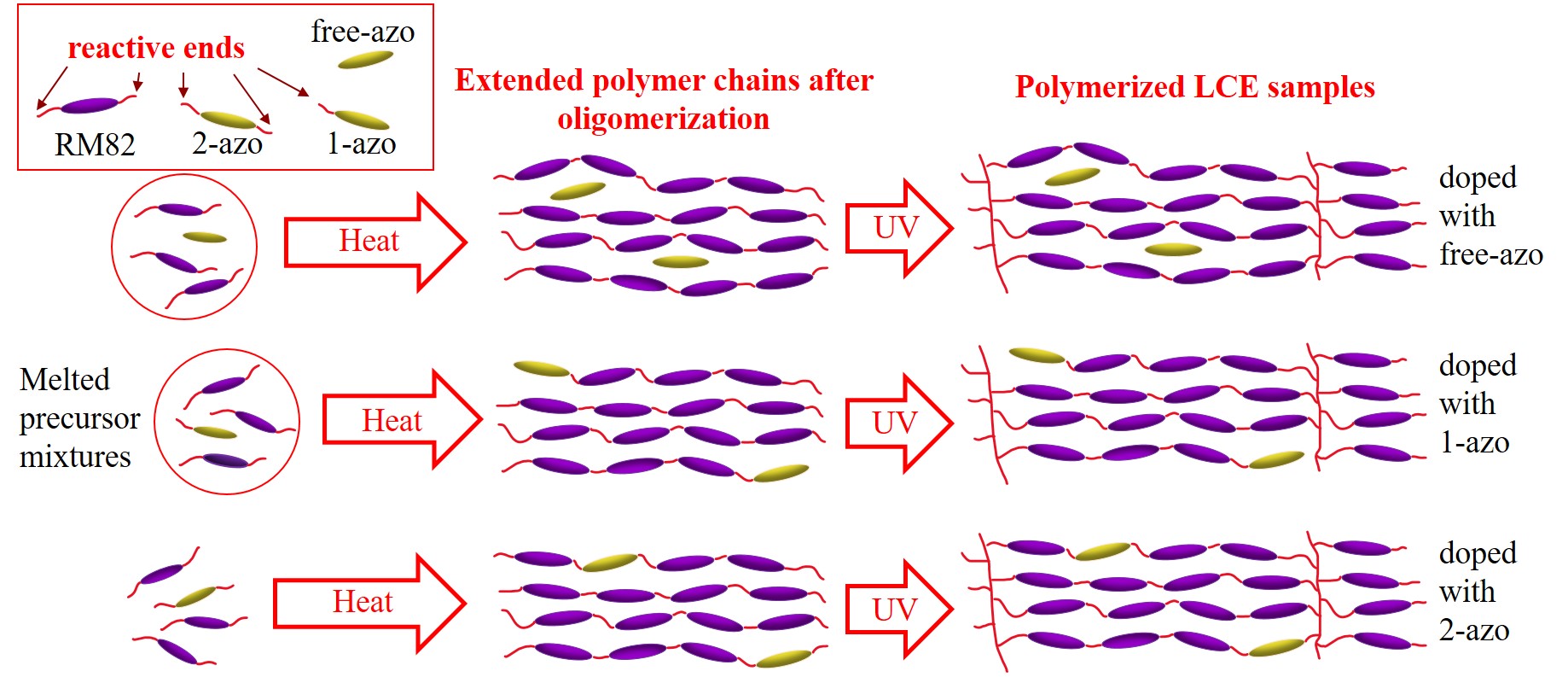}
\caption{Networks of azo-doped acrylate based LCEs}
\label{fig:network}
\end{figure}

\section{Experimental Results}

\subsection{Elastic Moduli}

\begin{figure}[h]
\centering
a.\includegraphics[width=0.3\linewidth]{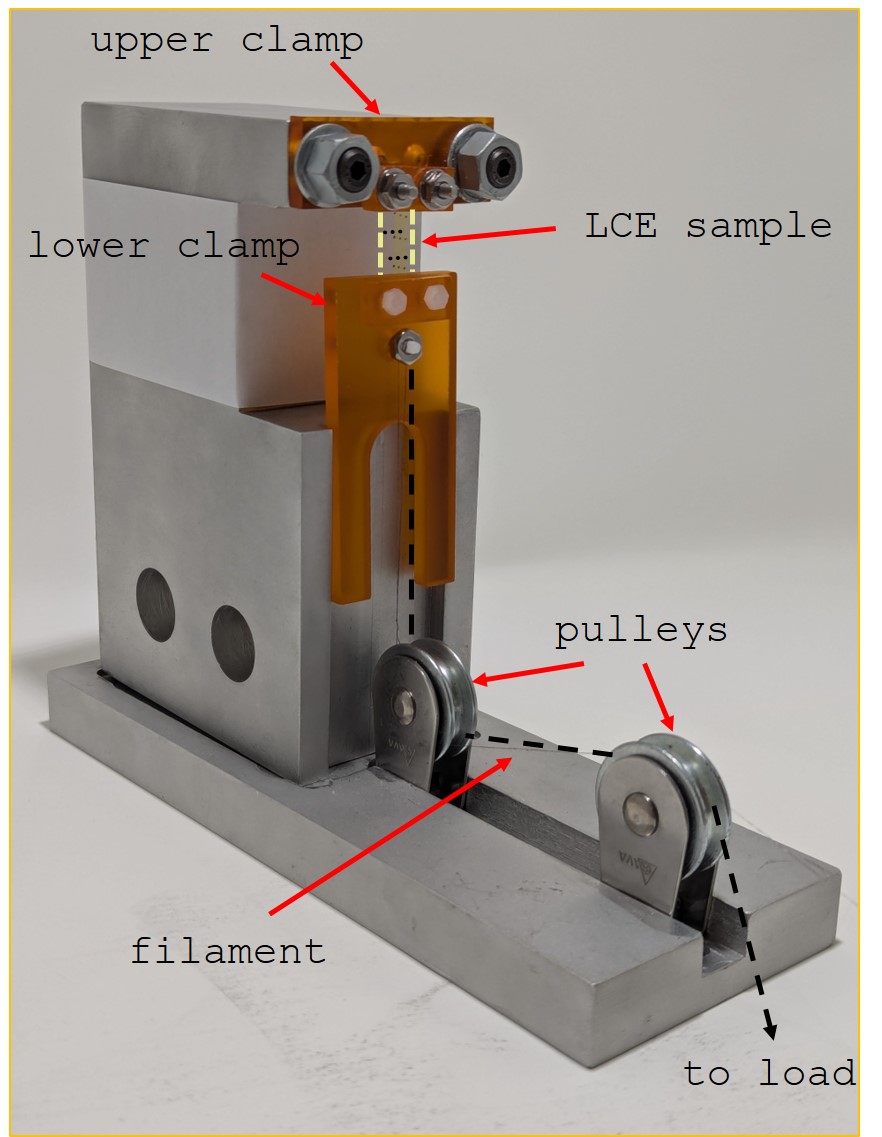}
b.\includegraphics[width=0.5\linewidth]{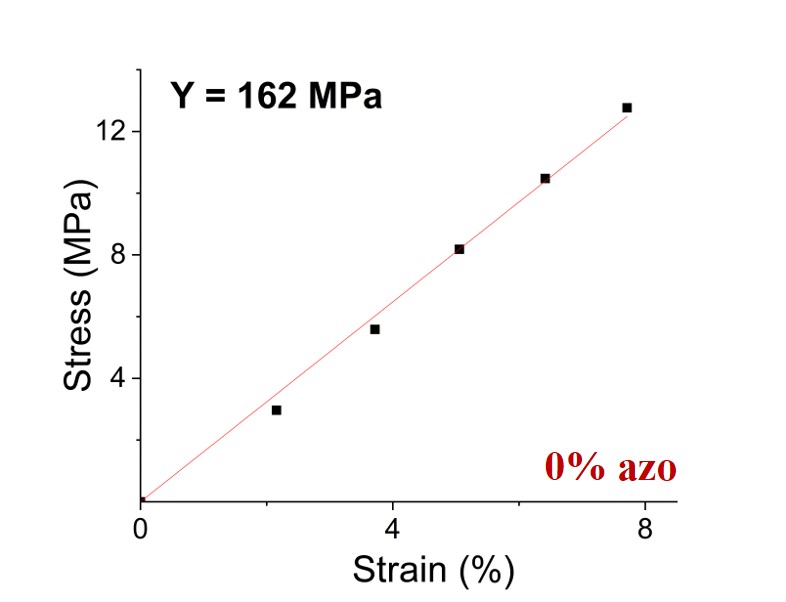}
\caption{a. Tensile force apparatus. b. Young's modulus data
for LCE sample with no azo dye.}
\label{fig:ImageAna}
\end{figure}

Young's modulus of all samples was measured using custom-built apparatus 
\cite{5 elastic}. For these measurements LCE samples were cut into $4~mm$ $%
\times $ $20~mm$ strips with the director oriented parallel to the long
edge. Samples marked with black dots were clamped at top and bottom and a
tensile force was applied. A photograph of the sample was taken for each
load and the strain was determined using ImageJ software \cite{5 elastic}. A
schematic of the device and experimental setup, together with typical data
is shown in Fig.~\ref{fig:ImageAna}.

\begin{figure}[h]
\centering
\includegraphics[width=1\linewidth]{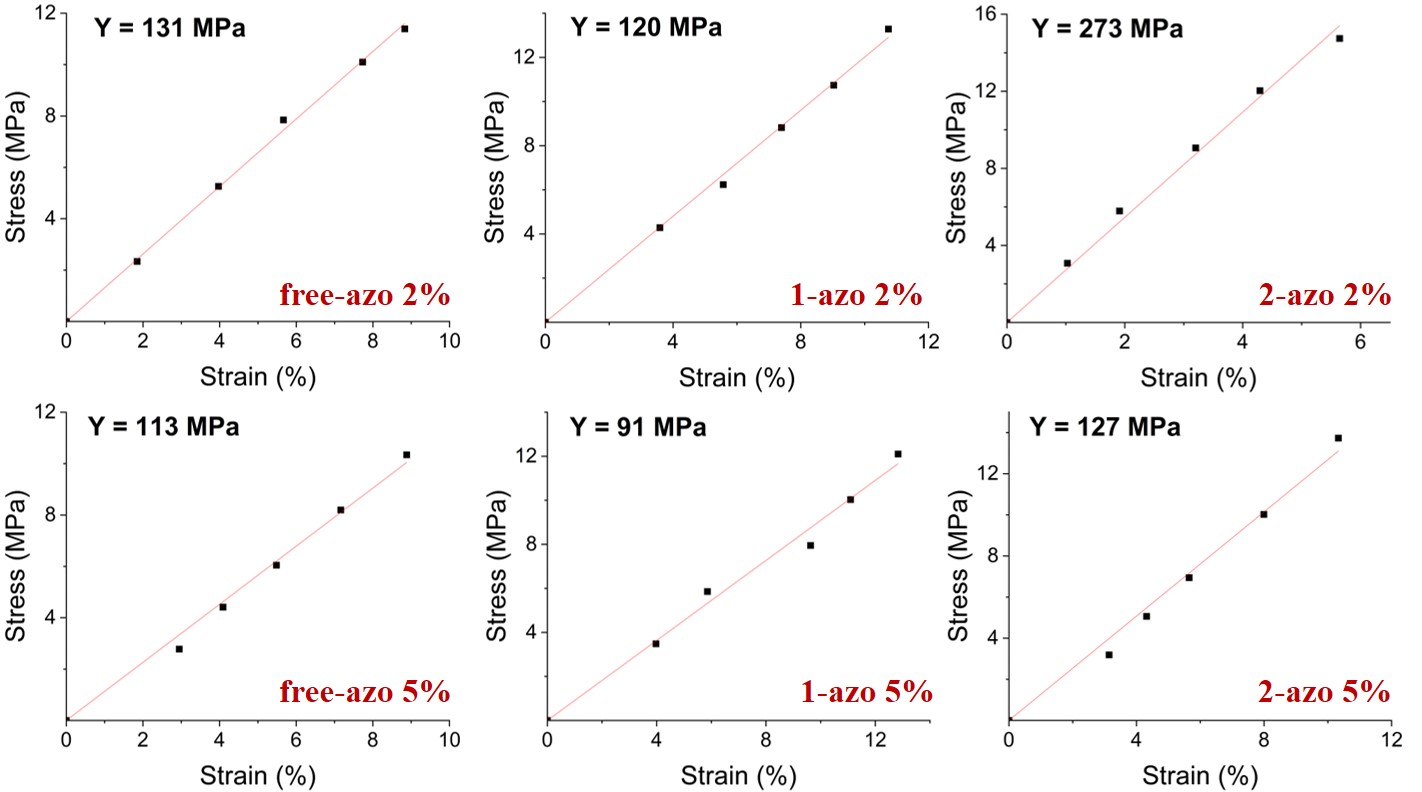}
\caption{Young's moduli data for all samples. Top and bottom rows are for 2\% and
5\% free-azo, 1-azo and 2-azo samples, respectively.}
\label{fig:Youngs}
\end{figure}

The LCE samples without any applied stress are not completely flat.
Elongating a sample initially therefore requires very little stress, which
then increases nonlinearly until the linear regime is reached. Measurements
are shown in the linear regime. Data and Young's moduli for all samples are
shown in Fig.~\ref{fig:Youngs}.

\subsection{Thermal Stress Measurements}

\begin{figure}[h]
\centering
\includegraphics[width=0.6\linewidth]{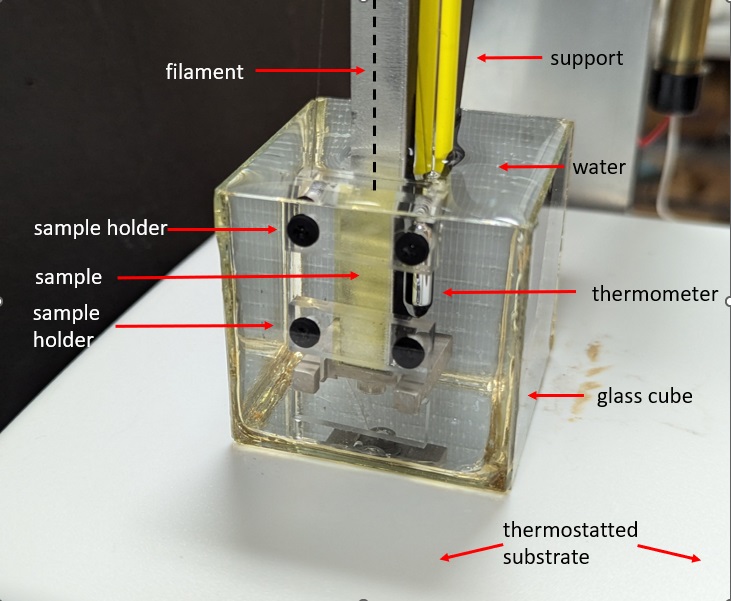}
\caption{Thermal and photostress setup. The filament location is
indicated by dashed line. Thermal insulation has been removed for
visibility. }
\label{fig:ExpSetUp}
\end{figure}

Our samples for thermal and photostress measurements were $30~mm$ long, $10~mm$ wide and $27-32~\mu m$ thick, with
nematic alignment along their long edge. They were held with the long edge
vertical and their two $10~mm$ edges clamped between two rigid horizontal
plexiglass struts leaving a $10$~$mm\times 10$~$mm$ sample area between holders. The
bottom edge was held in fixed position by a rigid support, while the top was
connected to an Entran force sensor, whose height could be adjusted, with a
nearly inextensible thin (Fireline jewelry thread) filament. The sample and
its holders were under water in a thermostatted container. An image of the
setup, used for both thermal and photostress measurements, is shown in Fig.~%
\ref{fig:ExpSetUp}.

Thermal stress was determined by the following procedure. The initial stress
on the sample was set by adjusting the height of the Entran sensor. The
water temperature was increased in steps. At each temperature, the Entran
voltage was recorded continuously for $1\min $, and was then averaged over
time to provide the force and then the stress exerted by the sample. The
thermal stress was obtained by subtracting the initial stress from the
measured stress. The heating rate between measurements was $0.25^{\circ
}\mathrm{C}/\min $.

Thermal stress data from azo-containing samples is presented below together
with photostress.

\subsection{Transmission spectra}

Photoactuation in these samples is due to photoisomerization of the azo
moiety. When azo isomers, predominantly in the extended \textit{trans}-
configuration, absorb light at $365~nm$, they undergo photoisomerization to
the more compact \textit{cis}- form, which has absorption peak near $450~nm$%
. Information about dynamics of the population distribution can be inferred
from transmission spectra.

\begin{figure}[h]
\centering
\includegraphics[width=1\linewidth]{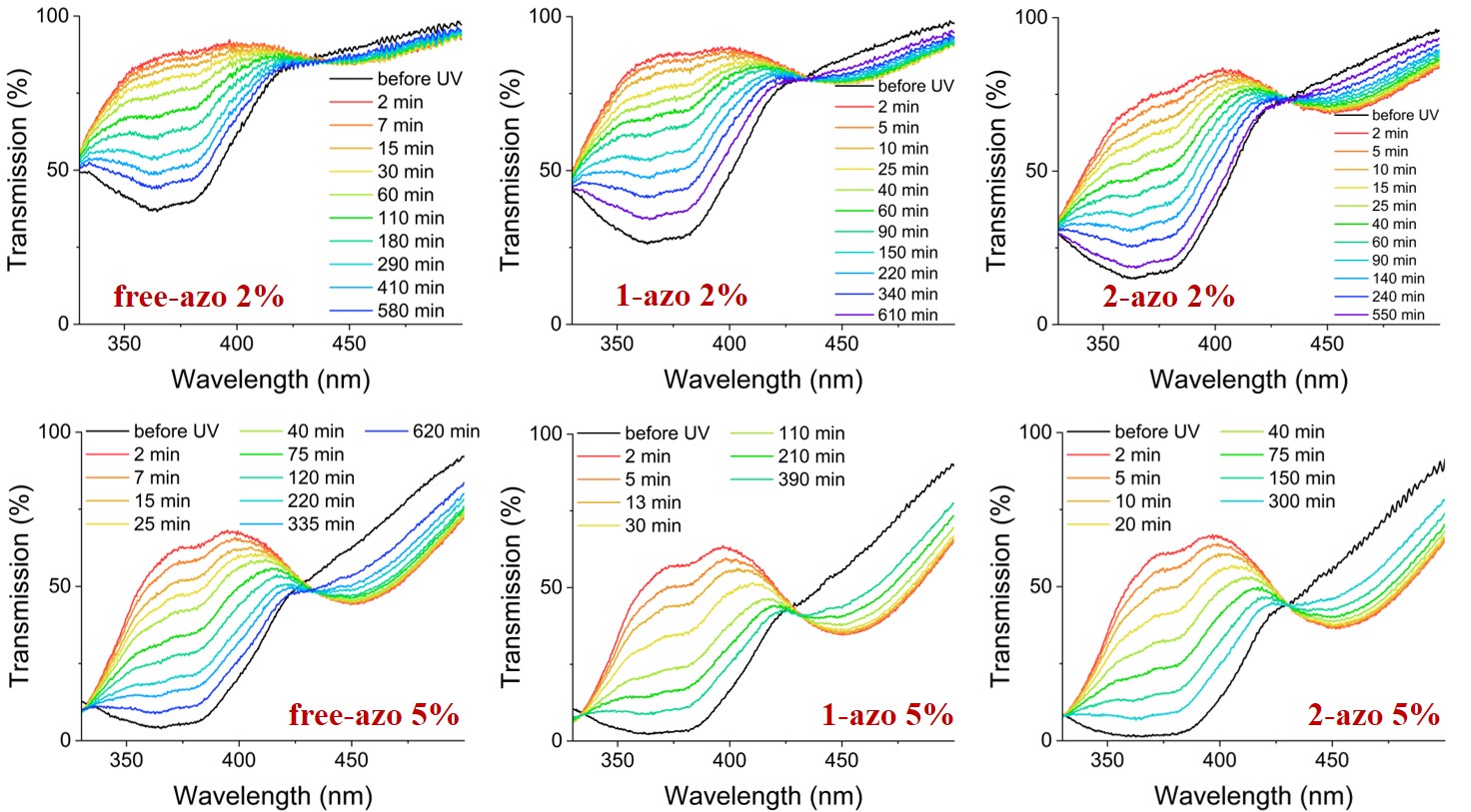}
\caption{Transmission spectra of all samples. Top and bottom rows are for
2\% and $5\%$ of free-azo, 1-azo and 2-azo samples, respectively.}
\label{fig:Spectra}
\end{figure}

We have illuminated our samples with light from a $500~mW$ LED at $365~nm$.
The intensity of the illumination was  $250$~$mW/cm^{2}$ with duration $10~s$%
. We then measured the transmittance spectra, whose minima give information
about the relative density of the \textit{trans}- and \textit{cis}- isomers,
as function of time. \ Transmission measurements were performed using an
HR4000CG-UV-NIR (Ocean Optics Inc., USA) spectrometer. Spectra were
normalized with respect to the sample with no azo dye. The transmittance
relaxation data is shown in Fig.~\ref{fig:Spectra}.

Material and optical parameters of the samples are given in Table \ref%
{table:Samples}. The absorption cross-section is $\sigma =1/(\rho
l_{d})$, where $\rho $ is the number density, and $l_{d}$ is the decay
length.

\begin{table}[tbp]
\caption{Sample material and optical parameters}
\label{table:Samples}\centering
\begin{tabular}{|c||c|c|c|c|c|c|}
\hline
Sample & Thickness & Max & Number & Decay length & Absorption & Cis-lifetime
\\ 
& ($\mu m$) & Absorbance & density of dye & ($\mu m)$ & cross-section & (s)
\\ 
&  & (Arb. units) & ($m^{-3}) $ &  & ($m^2$) &  \\ \hline
free-azo 2\% & 27 & 0.43 & $1.86\times 10^{25}$ & 63 & $8.5\times 10^{-22}$
& $1.2\times 10^4$ \\ \hline
1-azo 2\% & 27 & 0.57 & $1.94\times 10^{25}$ & 47 & $1.1\times 10^{-21}$ & $%
9.3\times 10^3$ \\ \hline
2-azo 2\% & 28 & 0.82 & $1.97\times 10^{25}$ & 34 & $1.49\times 10^{-21}$ & $%
4.4\times 10^3$ \\ \hline
free-azo 5\% & 31 & 1.36 & $4.7\times 10^{25}$ & 23 & $9.2\times 10^{-22}$ & 
$5.9\times 10^3$ \\ \hline
1-azo 5\% & 32 & 1.61 & $4.64\times 10^{25}$ & 20 & $1.08\times 10^{-21}$ & $%
6.5\times 10^3$ \\ \hline
2-azo 5\% & 27 & 1.87 & $4.84\times 10^{25}$ & 14 & $1.48\times 10^{-21}$ & $%
3.6\times 10^3 $ \\ \hline
\end{tabular}%
\end{table}

Although transmittance is clearly higher for the $2\%$ than for the $5\%$
azo containing samples, there is very little difference in the extent of
isomerization in the free-azo, 1-azo and 2-azo samples. We note here that
the isomerization process is only weakly dependent on temperature, given the
large excitation energy needed for azo \textit{trans-cis} isomerization \cite%
{regimes}.

\subsection{Photostress and Thermal Stress Measurements}

Photostress measurements were carried out using the same setup as thermal
stress measurements, shown in Fig.~\ref{fig:ExpSetUp}. The sample was
illuminated with UV\ light at $365~nm$\thinspace\ from a $500~mW$ LED
(Prizmatix Ltd.) at a distance of $6~cm$ from the sample. The light source
was unpolarized, so polarization dependent effects\cite{White1, Warner06}
need not be considered. The illuminated area of the sample was $10~mm\times
10~mm$. The illumination sequence consisted of: UV OFF for $10~s$, then,
repeatedly, UV ON for $200~ms$, UV OFF for $1~s$. Since the \textit{cis-}
thermal relaxation time is very slow, as indicated in Table \ref%
{table:Samples}, after each measurement the sample was illuminated with
light at $460~nm$ for $1~s$ to drive the \textit{cis-trans} isomerization.

\begin{figure}[h]
\centering
\includegraphics[width=1\linewidth]{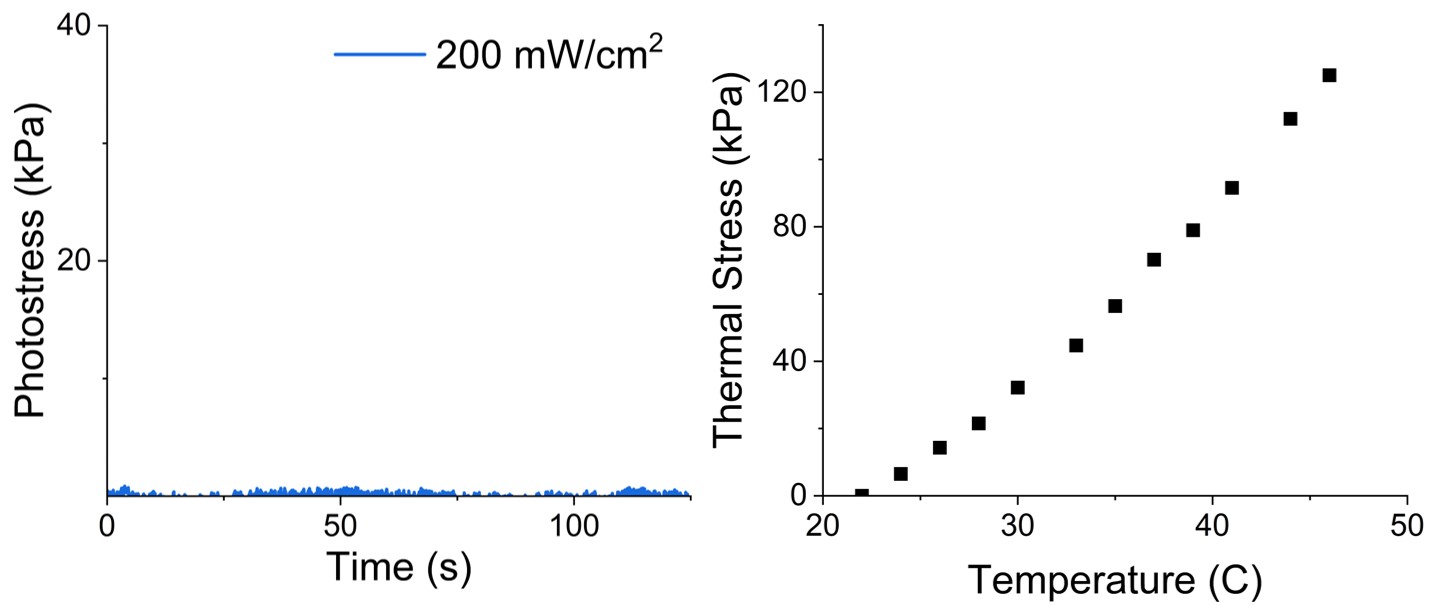}
\caption{Photostress and thermal stress in acrylate LCE samples not
containing azo dye. The initial stress is 50
kPa.}
\label{fig:Stress_0zao}
\end{figure}

To illustrate that the photomechanical effect is result of the presence of
azo dyes, we show the thermal and photoresponse of samples without azo dye
in Fig.~\ref{fig:Stress_0zao}. There is no observable photostress in our
thermostatted samples without azo dyes.

\begin{figure}[h]
\centering
\includegraphics[width=1\linewidth]{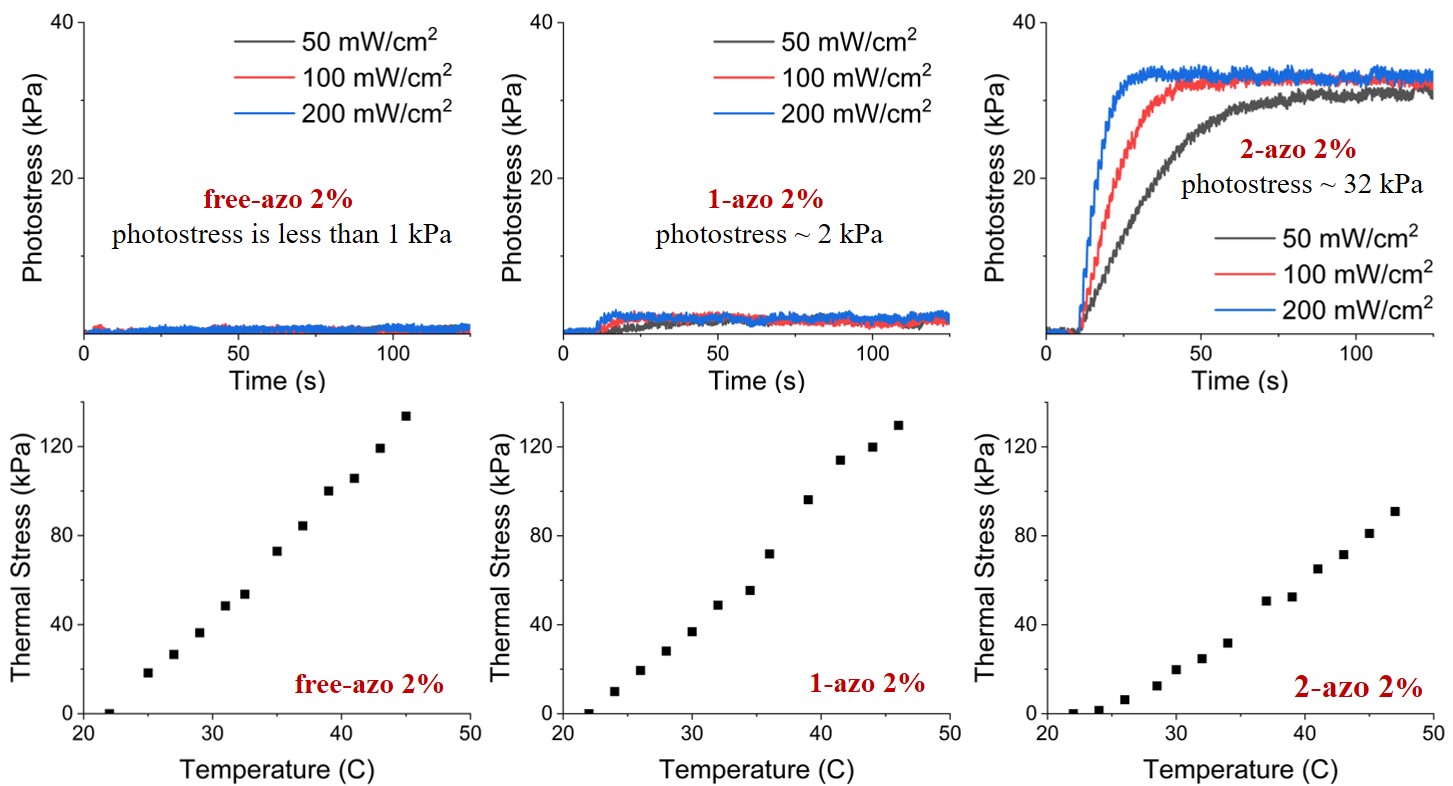}
\caption{Photostress (top) and thermal stress (bottom) in acrylate LCE
samples containing 2\% 0-azo, 1-azo, and 2-azo dye. The initial stress is 50
kPa.}
\label{fig:2percent}
\end{figure}
\begin{figure}[h]
\centering
\includegraphics[width=1\linewidth]{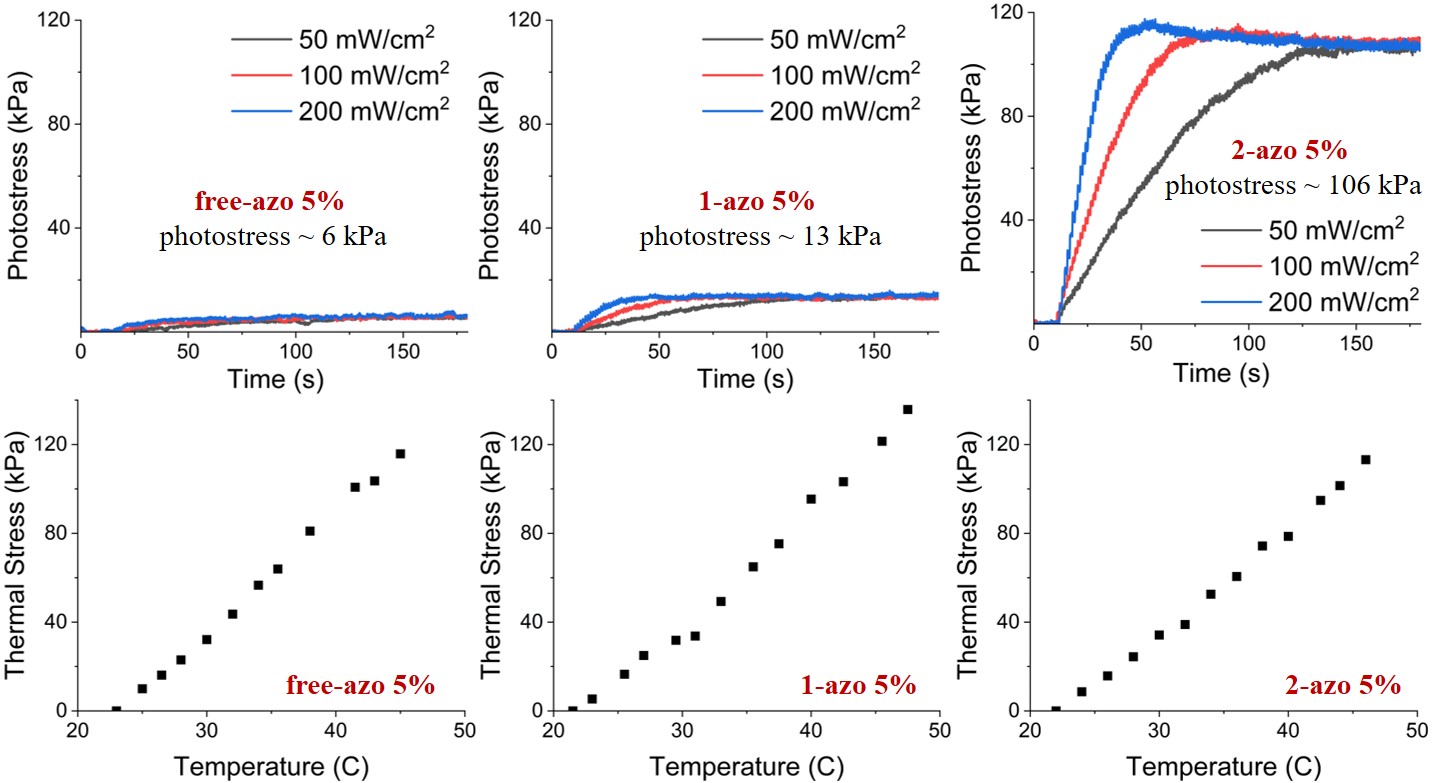}
\caption{Photostress (top) and thermal stress (bottom) in acrylate LCE
samples containing 5\% 0-azo, 1-azo, and 2-azo dye. The initial stress is 90
kPa.}
\label{fig:5percent}
\end{figure}

Photostress measurement results, together with those from thermal stress
measurements, are shown in Figs.~\ref{fig:2percent} and \ref{fig:5percent}.
The results of measurements clearly indicate that the thermal stresses are
essentially indentical for free-azo, 1-azo and 2-azo samples with the same
initial stresses, however, photostresses of 2-azo samples differ
dramatically from those of free-azo and 1-azo samples with the same inital
stresses for both 2\% and 5\% azo concentrations. Our interpretations of
these results are elucidated in the following discussion section.

\section{Discussion}

Heating changes the orientational order of nematic liquid crystals, which is
necessarily coupled to mechanical strain \cite{PGdG}, \cite{WarnerB}. This
coupling persists in solid LC elastomers \cite{ppm1} and even in LC glasses 
\cite{ppm2}. Photoisomerization changes the shape of the azo-containing
molecules. This gives rise to two competing mechanisms: 1. change in
orientational order due to the azo-containing molecules being less
liquid-crystal like (the shape is less elongated and the polarizability is
less anisotropic), and 2. a direct contractile stress due to the contraction
of the two ends of the aligned azo-mesogens due to photoisomerization.

The first of these, the change in orientational order parameter has often
been referred to as the major mechanism for photoactuation in LCEs  \cite%
{CviklinskiTerentjev02, White1, WarnerB, ppm3, Warner04,Warner06, Warner09}\
and has been seen as playing essentially the same role in both thermal and
photoactuation  \cite{ppm4, Warner04, Warner09}. The second, direct
contractile stress,  was already recognized by Finkelmann and Sanchez \cite%
{SanchezHeino11} who referred to the mechanism as `cooperative effect'.
Earlier studies have probed the effects of different azo dye attachments 
\cite{HoganTerentjev02, HarveyTerentjev07}, but the different structures of
the dyes used made comparisons difficult. Later work on siloxane elastomers 
\cite{SanchezHeino11} with two different kinds of 2-azo and 1-azo dyes (also
different from ours) indicated that higher photostress was exhibited by
2-azo samples. Although the justification is not clearly given in their
paper, it is suggested that mechanism 2 provides $60\%$ and mechanism 1 $40\%
$ of the stress.

Our three samples, with the 2-azo, 1-azo and free-azo architectures, exhibit
essentially identical thermal stress as shown in the bottom rows of Figs.~%
\ref{fig:2percent} and \ref{fig:5percent}. This implies that the differences
in network architecture do not affect the thermally induced change in order
parameter or the stress resulting from the order parameter change. The
thermal stress is due to the change of order parameter caused by a change of
temperature, and the stress is primarily the consequence of the coupling of
the order parameter and the stress.

The three samples, with the 2-azo. 1-azo and free-azo architectures exhibit
the same photoisomerization, as indicated by spectroscopic results shown in
Fig.~\ref{fig:Spectra}. If the photostress was primarily due to the change
in order parameter via mechanism 1, then the photostress would be
essentially the same for all samples, since the photoisomerization is
essentially the same. If the photostress was primarily due to a direct
contractile stress due to mechanism 2, then the result for the 2-azo sample
would be dramatically different from the response of the 1-azo and free-azo
samples. Since the difference in the observed photostress is dramatically
larger for the 2-azo samples than those for the 1-azo and free-azo samples,
we conclude that the primary mechanism is the contractile stress due to
photoisomerization in mechanism 2.

In light of the above, it is interesting to ask: what then is the role of
liquid crystallinity in the photoresponse of 2-azo samples? \ Likely the
role is to provide alignment of the azo\textit{\ }moiety when the sample is
prepared. Acrylate LCEs, with or without azo compounds, are essentially
aligned nematic samples, which would also align the predominantly \textit{%
trans- }azo isomers. \ Our samples produce uniaxial stress because the
stress-producing azo moiety is aligned, and the contraction of the
unidirectionally aligned azo-nematogens, with their covalent bonds at two
ends, efficiently transfer stress to the bulk network. It appears then that
the main role of liquid crystallinity in the 2-azo samples is to provide the
alignment of the stress-producing azo moities for the production of
unidirectional stress.

\section{Summary}

Our experimental results indicate that photostress created in 2-azo acrylate
LCEs is due to direct unidirectional contraction of the network caused by
shape change of the azo moiety during photoisomerization. Stress associated
with order parameter change also exists, as observed in 1-azo and free-azo
samples, but it is about one order of magnitude smaller than photosress in
2-azo materials. In the 2-azo samples, the traditional actuating mechanism
via order parameter change is therefore not significant; the primary role of
liquid crystallinity is simply to align the photoresponsive azo containing
molecules during network formation.

\section{Acknowledgements}

We are grateful to Professors Dirk Broer and Danqing Liu for numerous
enlightening discussions as well as some sample materials. We are indebted
to Dr. Bahman Taheri of AlphaMicron Inc. who kindly presented this work at
ILCC2024 on our behalf. We acknowledge support in part from the Office of
Naval Research through the MURI on Photomechanical Material Systems (ONR
N00014-18-1-2624).

\end{document}